\documentstyle[aps,prd]{revtex}
\begin{document}
\def\parslash{{\partial \!\!\! /}}

\title {Decoherence in Field Theory: General Couplings and Slow Quenches}

\author{F.\ C.\ Lombardo$^{1}$  \footnote{Electronic address: lombardo@df.uba.ar},
F.\ D.\ Mazzitelli $^1$ \footnote{Electronic address: fmazzi@df.uba.ar},
and R.\ J.\ Rivers $^2$ \footnote{Electronic address: r.rivers@ic.ac.uk}}

\address{{\it
$^1$ Departamento de F\'\i sica, Facultad de Ciencias Exactas y Naturales\\
Universidad de Buenos Aires - Ciudad Universitaria, Pabell\' on I\\
1428 Buenos Aires, Argentina\\
$^2$  Theoretical Physics Group, The Blackett Laboratory, \\ Imperial
College, London SW7 2BZ, U.K.}}

\maketitle

\begin{abstract}

We study the onset of a classical order parameter after a second-order
phase transition in quantum field theory.
We consider a quantum scalar field theory in which the system-field
(long-wavelength modes), interacts  with its environment,
represented both by a set of scalar fields and by its own short-wavelength
modes. We compute the decoherence times for the system-field modes and compare
them with the other time scales of the model. We analyze
different couplings between the system and the environment for slow
quenches. Within our approximations decoherence is in general a short time
event.
\end{abstract}



\section{Introduction}

The emergence of classical physics from quantum behaviour is
important for several physical phenomena in the early Universe.
This is beyond the fundamental requirement that only after the
Planck time can the metric of the Universe be assumed to be
classical. For example,
\begin{itemize}
\item
The  inflationary era is assumed to have been induced by scalar
inflaton fields, with simple potentials \cite{linde}. Such fields
are typically assumed to have classical behaviour, although in
principle a full quantum description should be used.
\item
The origin of large scale structure in the Universe can be traced
back to quantum fluctuations that,
after crossing the horizon, were frozen and became classical, stochastic,
inhomogeneities \cite{todosLSS}.
\item
It is generally assumed that several phase transitions have
occurred during the expansion of the Universe \cite{old}. As in
the case for the inflaton fields, the (scalar) order parameter
fields that describe these transitions are described classically.
However, the description of early universe phase transitions from
first principles is intrinsically quantum
mechanical\cite{cormier}.
\item
As a specific application \cite{Kibble} of the previous point, the
very notion of topological defects (e.g. strings and monopoles)
that characterize the domain structure after a finite-time
transition, and whose presence has consequences for the early
universe, is based on this assumption of classical behaviour for
the order parameter\cite{vilen}, as it distributes itself between
the several degenerate ground states of the ordered system.

\end{itemize}

In the present paper we are concerned with the third point above,
the quantum to classical transition of the order parameters in
second order phase transitions.  Any approach must take into
account both the quantum nature of the order parameter and the
non-equilibrium aspects of the process. The problem of the quantum
to classical transition in the context of inflationary models was
first addressed by Guth and Pi \cite{guthpi}. In that work, the
authors used an inverted harmonic oscillator as a toy model to
describe the early time evolution of the inflaton, starting from a
Gaussian quantum state centered on the maximum of the potential.
They subsequently showed that, according to Schr\"odinger's
equation, the initial wave packet maintains its Gaussian shape
(due to the linearity of the model). Since the wave function is
Gaussian, the Wigner function is positive for all times. Moreover,
it peaks on the classical trajectories in phase space as the wave
function spreads. The Wigner function can then be interpreted as a
classical probability distribution for coordinates and momenta,
showing sharp classical correlations at long times. In other
words, the initial Gaussian state becomes highly squeezed and
indistinguishable from a classical stochastic process. In this
sense, one recovers a classical evolution of the inflaton rolling
down the hill.

A similar approach has been used by many authors to describe the
appearance of classical inhomogeneities from quantum fluctuations
in the inflationary era \cite{staro}. Indeed, the Fourier modes of
a massless free field in an expanding universe satisfy the linear
equation
\begin{equation}
\phi_k''+(k^2-{a''\over a})\phi_k = 0.
\end{equation}
For sufficiently long-wavelengths ($k^2\ll a''/a$), this equation
describes an unstable oscillator.
If one considers an initial Gaussian wave function, it will remain
Gaussian for all times, and it will spread with time. As with the
toy model of Guth and Pi, one can show that classical correlations
do appear, and that the Wigner function can again be interpreted
as a classical probability distribution in phase space. [It is
interesting to note that a similar mechanism can be invoked to
explain the origin of a classical, cosmological magnetic field
from amplification of quantum fluctuations].

However, classical correlations are only one aspect of classical
behaviour. It was subsequently recognized that, in order to have a
complete classical limit, the role of the environment is crucial,
since its interaction with the system distinguishes the field
basis as the pointer basis \cite{kiefer}. [We are reminded that,
even for the fundamental problem of the space-time metric becoming
classical, simple arguments based on minisuperspace models suggest
that the classical treatment is only correct because of the
interaction of the metric with other quantum degrees of freedom
\cite{halli}.]

While these linear instabilities cited above characterise   {\it
free} fields, the approach fails when interactions are taken into
account. Indeed, as shown again in simple quantum mechanical
models (e.g. the anharmonic inverted oscillator), an initially
Gaussian wave function becomes non-Gaussian when evolved
numerically with the Schr\"odinger equation. The Wigner function
now develops negative parts, and its interpretation as a classical
probability breaks down \cite{diana}. One can always force the
Gaussianity of the wave function by using a Gaussian variational
wave function as an approximate solution of the Schr\"odinger
equation, but this approximation deviates significantly from the
exact solution as the wave function probes the non-linearities of
the potential \cite{diana,stancioff}.

When interactions are taken into account, classical behaviour is
recovered only for "open systems", in which the relevant 
degrees of freedom interact with their environment. When this
interaction produces {\it both} a diagonalization of the reduced
density matrix and a positive Wigner function, the quantum to
classical transition is completed \cite{giulinibook}.

Going from quantum mechanical toy models to  quantum field theory
is, of course, extremely difficult. For this reason, several
authors \cite{mihaila} have considered different approximations in
quantum mechanics, and compared them to the exact results. If
successful in quantum mechanics, they could be implemented with
greater confidence in field theory. As already mentioned, when
this procedure is applied in the context of  {\it closed systems},
the conclusion is that, in general, mean-field approximations  do
not reproduce the evolution of the system at late
time\cite{diana,stancioff,mihaila}. Therefore, in principle, there
is not reason to believe they will do so in field theory. In spite
of this, computational necessity has lead many authors to perform
calculations in closed field theories with Hartree, mean field, or
$1/N$ approximations, since they are non-perturbative,
well-defined and suitable for numerical calculations
\cite{salman,mottola,devega,Boya}. In such calculations, classical
correlations do appear in some field theory models
\cite{mottola,devega}. However, since such decoherence (in a
time-averaged sense) takes place at long times after the
transition has been achieved initially, when the mean field
approximation has broken down, this may be an artifact of the
Gaussian-like approximations \cite{diana}.

In a previous paper \cite{diana} we considered similar arguments
for quantum mechanics, but for {\it open systems}. When an
anharmonic inverted oscillator is coupled to a high temperature
environment, it becomes classical very quickly, even before the
wave function probes the non-linearities of the potential.  Being
an early time event, the quantum to classical transition can now
be studied perturbatively. In general, recoherence effects are not
expected \cite{nuno}. Taking these facts into account, we have
extended the approach to field theory models \cite{lomplb}. In
field theory, one is usually interested in the long-wavelengths of
the order parameter. Moreover, the early universe is replete with
fields of all sorts which comprise a rich environment. For this
reason, we considered a model in which the order parameter
interacts with a large number of environmental fields, including
its own short-wavelengths. Assuming weak coupling and high
critical temperature, we have shown  that decoherence is a short
time event, shorter than the spinodal time $t_{\rm sp}$, which is
essentially that time by which the order parameter field has
sampled the degenerate ground states. As a result, perturbative
calculations are justified\cite{lomplb}. Subsequent dynamics can
be described by a stochastic Langevin equation, the details of
which are only known for early times.

Our approach in Ref.\cite{lomplb} has some connections with
well-established classical behaviour of thermal scalar field
theory \cite{htl} at high temperature. In many articles it has
been shown that, at high temperatures, the behaviour of
long-wavelength modes is determined by classical statistical field
theory. The effective classical theory is obtained after
integrating out the hard modes with $k\geq gT$. Although similar,
this approach has some important differences from ours: the
``classical behaviour'' in this soft thermal mode analysis is
defined through the coincidence of the quantum and the statistical
correlation functions. In particular, thermal equilibrium is
assumed to hold at all times. Finally, the cutoff that divides
system and environment depends on the temperature, which is
externally fixed. In our approach, the quantum to classical
transition is defined by the diagonalization of the reduced
density matrix, which is not assumed to be thermal. In phase
transitions the separation between long and short-wavelengths is
determined by their stability, which depends on  the parameters of
the potential.

In Ref.\cite{lomplb} we only considered the case of an
instantaneous quench, and bi-quadratic coupling between the system
and the environment. In Ref.\cite{lomplb2} we began to extend
those results to the case of a slow quench. In this paper we
provide the details of that analysis and extend it to other
couplings between system and environment. The consideration of
slow quenches is very important since the Kibble-Zurek mechanism
predicts the relation between the subsequent domain structure and
the quench time \cite{kzm,Rivers2,laguna,zurek3} (by indirectly
counting defects).

The paper is organized as follows. In Section 2 we introduce our
models. These are theories containing a real system field $\phi$,
which undergoes a transition, coupled to other scalar fields
$\chi_{\rm a}$ (${\rm a} = 1,2,...,N$), which constitute the
external part of the environment. Gauge fields bring their own
specific difficulties and we shall not discuss them here. We
compute the influence functional by integrating out the
environmental fields for different couplings. Section 3 is
dedicated to reviewing the evaluation of the master equation and
the diffusion coefficients which are relevant in order to study
decoherence. In Section 4 we evaluate upper bounds on the
decoherence time for slow quenches. As we will see, provided
quenches are not to slow, decoherence takes place before the field
samples the minimum of the potential, i.e. decoherence time is
typically shorter than the spinodal time. However, since all
relevant timescales depend only logarithmically upon the
parameters of the theory, it is necessary to keep track of $O(1)$
prefactors carefully, something that we rather took for granted in
Ref.\cite{lomplb} and in Ref.\cite{lomplb2}. We will also show
that the bi-quadratic coupling is the most relevant for the
quantum to classical transition. Section 5 contains the
conclusions of our work. Two short appendices fill in some of the
detail.

\section{The Influence Action for an External
Environment} \label{sec:model}

For the infinite degree of freedom quantum field $\phi$ undergoing
a continuous transition, the field ordering after the transition
begins is due to the growth in amplitude of its unstable
long-wavelength modes. For these modes the environment consists of
the short-wavelength modes of the field, together with all the
other fields with which  $\phi$ inevitably
interacts\cite{huzhang,lombmazz,greiner} in the absence of
selection rules. The inclusion of explicit environment fields is
both a reflection of the fact that a scalar field in isolation is
physically unrealistic, as well as providing us with a systematic
approximation scheme.

The $\phi$-field describes the scalar order parameter, whose
${\cal Z}_2$ symmetry is broken by a double-well potential.
Specifically, we take the simplest classical action with scalar
and environmental fields $\chi_{\rm a}$
\begin{equation}
S[\phi , \chi ] = S_{\rm syst}[\phi ] + S_{\rm env}[\chi ] +
S_{\rm int}[\chi_{\rm a},\phi ], \label{action0}
\end{equation}
where (with $\mu^2$, $m^2 >0$ )
\begin{equation}
S_{\rm syst}[\phi ] = \int d^4x\left\{ {1\over{2}}\partial_{\mu}
\phi\partial^{\mu} \phi + {1\over{2}}\mu^2 \phi^2 -
{\lambda\over{4!}}\phi^4\right\}, \nonumber \end{equation}
\begin{equation}
S_{\rm env}[\chi_{\rm a} ] = \sum_{\rm a=1}^N\int d^4x\left\{
{1\over{2}}\partial_{\mu}\chi_{\rm a}
\partial^{\mu}
\chi_{\rm a} - {1\over{2}} m_{\rm a}^2 \chi^2_{\rm a}\right\}.
\nonumber
\end{equation}

The most important interactions will turn out to be of the
biquadratic form
\begin{equation}
S_{\rm int}[\chi_{\rm a},\phi ] = S_{\rm qu}[\phi ,\chi ] = -
\sum_{\rm a=1}^N\frac{g_{\rm a}}{8} \int d^4x ~ \phi^{\rm 2} (x)
\chi^{\rm 2}_{\rm a} (x). \label{Sint}
\end{equation}
Even if there were no external $\chi$ fields with a quadratic
interaction of kind (\ref{Sint}), the interaction between short
and long-wavelength modes of the $\phi$-field can be recast, in
part, in this form (see later), showing that such a term is
obligatory.  The generalization to a complex field $\phi$ is
straightforward, and has been considered elsewhere \cite{lomplb2}.

Later, we shall consider additional interactions to the
biquadratic interaction of a bilinear form
\begin{equation} S_{\rm bilin}[\phi ,\chi ] = - \sum_{\rm
a=1}^N\frac{g'_{\rm a}\mu}{4} \int d^4x ~ \phi^{\rm 2} (x)
\chi_{\rm a} (x), \label{Sint2}
\end{equation} in which the environment couples linearly.

We could also have included fermionic Yukawa interactions but
these provide a smaller contribution to the diffusion constant
because of Fermi statistics (which gives a diffusion coefficient
relatively $O(\mu^2/T^{2})$). However, because the biquadratic
self-interaction is the overwhelming term in the diffusion
coefficient, in general we are not interested in effects that are
overshadowed by it, and we shall not consider Yukawa interactions
here.

Our exception to this rule of ignoring small contributions is in
the the inclusion of linear couplings of the form
\begin{equation}
S_{\rm lin}[\phi ,\chi_{\rm a} ] = - \sum_{\rm
a=1}^N\frac{g''_{\rm a}\mu^2}{4} \int d^4x ~ \phi (x) \chi_{\rm a}
(x), \label{Sint3}
\end{equation}
in which the order parameter field couples linearly to a linear
environment. Much if not most of the work on decoherence has been
for linear coupling to the environment of this type (e.g. see
\cite{unruh,kim}). While it is sensible in quantum mechanics, in
the context of quantum field theory such linear couplings signal
an inappropriate field diagonalisation, although they are exactly
solvable in some circumstances \cite{kim}.

Although the system field $\phi$ can never avoid the decohering
environment of its own short-wavelength modes, to demonstrate the
effect of an environment we first consider the case in which the
environment is taken to be composed only of the fields $\chi_{\rm
a}$. We are helped in this by the fact that environments have a
cumulative effect on the onset of classical behaviour. That is,
the inclusion of a further component of the environment {\it
reduces} the time $t_D$ it takes for the system to behave
classically (for the definition of $t_D$ see later). Thus it makes
sense to include the environment one part after another, since we
can derive an {\it upper} bound on $t_D$ at each step. The
short-wavelength modes of the $\phi$ field will be considered
last.

To keep our calculations tractable, we need a significant part of
the environment to have a strong impact upon the system-field, but
not vice-versa, from which we can bound $t_D$. The simplest way to
implement this is to take a large number $N\gg 1$ of scalar
$\chi_{\rm a}$ fields with comparable masses $m_{\rm a}\simeq \mu$
weakly coupled to the $\phi$, with $\lambda$, $g_{\rm a},g'_{\rm
a},g''_{\rm a}\ll 1$. Thus, at any step, there are $N$ weakly
coupled environmental fields influencing the system field, but
only one weakly self-coupled system field to back-react upon the
explicit environment.

We first consider the case in which the fields $\chi_{\rm a}$
interact through the biquadratic interaction (\ref{Sint}) alone.
For one-loop consistency at second order in our calculation of the
diffusion coefficient (that enforces classicality) it is
sufficient, at order of magnitude level, to take identical $g_{\rm
a} = g/\sqrt{N}$.  Further, at the same order of magnitude level,
we take $g\simeq\lambda$. This is very different from the more
usual large-N $O(N+1)$-invariant theory with one $\phi$-field and
$N$ $\chi_{\rm a}$ fields, dominated by the $O(1/N)$ $(\chi^2)^2$
interactions, that has been the standard way to proceed for a {\it
closed} system. With our choice there are no direct $\chi^4$
interactions, and the indirect ones, mediated by $\phi$ loops, are
depressed by a factor $g/\sqrt{N}$. In this way the effect of the
external environment qualitatively matches the effect of the
internal environment provided by the short-wavelength modes of the
$\phi$-field, but in a more calculable way.

For small $g$ the model has a continuous transition at a
temperature $T_{\rm c}$. The environmental fields $\chi_{\rm a}$
reduce $T_{\rm c}$ and, in order that $T_{\rm
c}^2/\mu^2=24/(\lambda + \sum g_{\rm a})\gg 1$, we must take
$\lambda + \sum g_{\rm a} \ll 1$, whereby $1\gg 1/\sqrt N\gg g$.
Further, with this choice the dominant hard loop contribution of
the $\phi$-field to the $\chi_{\rm a}$ thermal masses is
\[
\delta m^2_T = O (g T^2_{\rm c}/\sqrt{ N}) = O(\mu^2/N)\ll \mu^2.
\]
Similarly, the two-loop (setting sun) diagram which is the first
to contribute to the discontinuity of the $\chi$-field propagator
is of magnitude
\[
g^2 T_{\rm c}^2/N  = O(g\mu^2/N^{3/2})\ll\delta m^2_T, \] in turn.
That is, the effect of the thermal bath on the propagation of the
environmental $\chi$-fields is ignorable. This was our intention;
to construct an environment that reacted on the system field, but
was not reacted upon by it to any significant extent. In
particular, the infinite $N$ limit does not exist. Dependence on
$N$ is implicit through $T_{\rm c}$ as well as through the
couplings, for initial temperatures $T_0 = O(T_{\rm c})$. With
$\eta = \sqrt{6\mu^2/\lambda}$ determining the position of the
minima of the potential and the final value of the order
parameter, this choice of coupling and environments gives the
hierarchy of scales necessary for establishing a reliable
approximation scheme, as has been shown in \cite{lomplb}.

We shall assume that the initial states of the system and
environment are both thermal, at a high temperature
$T_{0}>T_{\rm c}$. We then imagine a change in the global
environment (e.g. expansion in the early universe) that can be
characterised by a change in temperature from $T_{0}$ to
$T_{\rm f}<T_{\rm c}$. That is, we do not attribute the
transition to the effects of the environment-fields.

Given our thermal initial conditions it is not the case that the
full density matrix has $\phi$ and $\chi$ fields
uncorrelated initially, since it is the interactions between them
that leads to the restoration of symmetry at high temperatures. Rather,
on incorporating the hard thermal loop 'tadpole' diagrams of the $\chi$
(and $\phi$) fields in the $\phi$ mass term leads to the effective
action for $\phi$ quasiparticles,
\begin{equation}
S^{\rm eff}_{\rm syst}[\phi ] = \int d^4x\left\{ {1\over{2}}\partial_{\mu}
\phi\partial^{\mu} \phi -
{1\over{2}} m_{\phi}^2(T_0) \phi^2 - {\lambda\over{4!}}\phi^4\right\} \label{Stherm}
\end{equation}
where $m_{\phi}^2(T_0)\propto (1-T_0/T_{\rm c})$ for $T\approx T_{\rm c}$. As
a result, we can
take an initial
factorised density matrix at temperature $T_0$ of the form ${\hat\rho}[T_0] =
{\hat\rho}_{\phi}[T_0] \times {\hat\rho}_{\chi}[T_0]$, where ${\hat\rho}_{\phi}[T_0]$
is determined by the quadratic part of $S^{\rm eff}_{\rm syst}[\phi ]$ and
${\hat\rho}_{\chi}[T_0]$ by
$S_{\rm env}[\chi_{\rm a} ]$. That is, the many $\chi_{\rm a}$ fields have
a large effect on $\phi$,
but the $\phi$-field has negligible effect on the $\chi_{\rm a}$.

Provided the change in temperature is not too slow the exponential
instabilities of the $\phi$-field grow so fast that the field has
populated the degenerate vacua well before the temperature has
dropped significantly below $T_{\rm c}$. Since the temperature
$T_{\rm c}$ has no particular significance for the environment
fields, for these early times we can keep the temperature of the
environment fixed at $T_{\chi}\approx T_{\rm c}$ (our calculations
are only at the level of orders of magnitude).  Meanwhile, for
simplicity the $\chi_{\rm a}$ masses are fixed at the common value
$ m\simeq\mu$.

Our interests in the transitions in the early universe (domain
structure, defect formation and, indirectly,
structure formation) and the subsequent field evolution
\cite{vilen} are naturally couched in terms of fields,
rather than particles. Since we need to be able to distinguish
between different classical system-field configurations
evolving after the transition, we will only be interested in the
field-configuration basis for this reduced density
matrix (in analogy with the usual quantum Brownian motion
model, e.g. \cite{zurekTA}). The resulting reduced density
matrix $\rho_{{\rm r}}[\phi^+,\phi^-,t]=\langle\phi^+
\vert {\hat\rho}_{\rm r} (t)\vert
\phi^- \rangle$ describes
the evolution of the system under the influence of the environment, and
is defined by
\[
\rho_{{\rm r}}[\phi^+,\phi^-,t] = \int \prod_{{\rm a}=1}^N {\cal D} \chi_{\rm a} ~
\rho[\phi^+,\chi_{\rm a} ,\phi^-,\chi_{\rm a} ,t],
\]
where $\rho[\phi^+,\chi^+_{\rm a} ,\phi^-,\chi^-_{\rm a},t]=
\langle\phi^+ \chi^+_{\rm a}\vert {\hat\rho}(t) \vert \phi^-
\chi^-_{\rm a}\rangle$ is the full density matrix. For the reasons
given above, the environment will have had the effect of making
the system effectively classical once $\rho_{\rm r}(t)$ is,
approximately, diagonal. Quantum interference can then be ignored
and we obtain a classical probability distribution from the
diagonal part of $\rho_{\rm r}(t)$, or equivalently, by means of
the reduced Wigner functional, which is positive definite after
$t_D$ \cite{diana}. For weak coupling there will be no
'recoherence' at later times in which the sense of classical
probability will be lost \cite{nuno}.

The temporal evolution of $\rho_{\rm r}$ is given by \[\rho_{\rm r}[\phi_{{\rm
f}}^+,\phi_{{\rm f}}^-,t]= \int d\phi_{{\rm i}}^+ \int d\phi_{{\rm i}}^- J_{\rm r}
[\phi_{{\rm f}}^+,\phi_{{\rm f}}^-,t\vert \phi_{{\rm i}}^+,\phi_{{\rm i}}^-,t_0] ~~
\rho_{\rm r}[\phi_{{\rm i}}^+ \phi_{{\rm i}}^-,t_0],
\]
where $J_{\rm r}$ is the reduced evolution operator

\begin{equation}
J_{\rm r}[\phi_{{\rm f}}^+,\phi_{{\rm f}}^-,t\vert \phi_{{\rm i}}^+,
\phi_{{\rm i}}^-,t_0] =
\int_{\phi_{{\rm i}}^+}^{\phi_{{\rm f}}^+} {\cal D}\phi^+
\int_{\phi_{{\rm i}}^-}^{\phi_{{\rm f}}^-}
{\cal D}\phi^- ~e^{i\{S[\phi^+] - S[\phi^-]\}} F[\phi^+,\phi^-].
\label{evolred}
\end{equation}
The Feynman-Vernon \cite{feynver} influence functional (IF)
$F[\phi^+, \phi^-]$ in (\ref{evolred}) can be written in terms of
influence action $\delta A[\phi^+,\phi^-]$ and the coarse grained
effective action (CGEA) $A[\phi^+,\phi^-]$ by (see \cite{lombmazz}
for formal definitions)
\begin{equation}
F[\phi^+,\phi^-] = \exp {i
\delta A[\phi^+,\phi^-]},\label{IA}
\end{equation}
\begin{equation}A[\phi^+,\phi^-] = S[\phi^+] - S[\phi^-] + \delta A[\phi^+,\phi^-].
\label{CTPEA}
\end{equation}

Beginning from the initial distribution, peaked around $\phi = 0$, we follow the evolution
of the system under
the influence of the environment fields. We will calculate the influence action to lowest
non-trivial order (two vertices) for large $N$. Similarly to the biquadratic interaction,
we assume weak coupling
$\lambda \simeq g'\simeq g''\ll 1$, where we have defined $g', g''$ by the order of magnitude
relations $g'_{\rm a}= g'/\sqrt{N}$ and $g''_{\rm a}= g''/\sqrt{N}$ respectively.

As we are considering weak coupling with the environment fields, we may expand the influence
functional
$F[\phi^+, \phi^-]$ up to second non-trivial order in coupling strengths. The general form of
the influence
action is then \cite{lombmazz,calhumaz} %
\begin{eqnarray}
\delta A[\phi^+,\phi^-] = &&\{\langle
S_{\rm int}[\phi^+,\chi^+_{\rm a}]\rangle_0 - \langle S_{\rm int}[\phi^-,\chi^-_{\rm a}]
\rangle_0\}
\nonumber
\\ &&+{i\over{2}}\{\langle S_{\rm int}^2[\phi^+,\chi^+_{\rm a}]\rangle_0 - \big[\langle
S_{\rm int}[\phi^+,\chi^+_{\rm a}]\rangle_0\big]^2\}\nonumber
\\ &&- i\{\langle S_{\rm int}[\phi^+,\chi^+_{\rm a}] S_{\rm int}[\phi^-,\chi^- _{\rm a}]
\rangle_0 -
\langle S_{\rm int}[\phi^+,\chi^+_{\rm a}]\rangle_0\langle S_{\rm int}[\phi^-,
\chi^-_{\rm a}]\rangle_0\}
\label{inflac} \\ &&+{i\over{2}}\{\langle S^2_{\rm int}[\phi^-,\chi^-_{\rm a}]\rangle_0
- \big[\langle
S_{\rm int}[\phi^-,\chi^-_{\rm a}]\rangle_0\big]^2\}.\nonumber \end{eqnarray}

For weak couplings it is relatively easy to compute the upper bound on the decoherence time
due to the interactions
$S_{\rm int}[\phi ,\chi ]$ of (\ref{Sint}), (\ref{Sint2}) and (\ref{Sint3}). We shall find that,
in our particular model, it is in general shorter than the spinodal time $t_{\rm sp}$, defined
as the time for which
\begin{equation}
\langle \phi^2\rangle_{t_{\rm sp}} \sim \eta^2= 6\mu^2/\lambda\,\,\,
.\label{deftsp}
\end{equation}
[More exactly, $t_{\rm sp}$ should be defined as the time that the
r.m.s of the field takes to reach the spinodal point, the point of
inflection in the potential, but the difference is only
logarithmically small, so we stay with (\ref{deftsp}).]

It is for this reason that, with the qualifications below, we can
use perturbation theory. In consequence, by the time that the
field is ordered it can be taken to be classical. This has
implications\cite{lomplb2} for the formation of the defects that
are a necessary byproduct of transitions.

\section{Master Equations and the Diffusion Coefficients} \label{sec:diff}

In this Section we will obtain the evolution equation for the
reduced density matrix (master equation), paying particular
attention to the diffusion term, which is responsible for
decoherence. We will closely follow the quantum Brownian motion
(QBM) example \cite{unruh,qbm}, translated into quantum field
theory \cite{lomplb,lombmazz}.

This Section contains a generalization of the result already given
in \cite{lomplb}, adapted to the slow quenches that are physically
relevant, in which $m^{2}_{\phi}(T)$ varies linearly in time. The
formal deduction of the master equation is not altered by the
duration of the quench. For the details we will follow
Ref.\cite{lombmazz}. The first step in the evaluation of the
master equation is the calculation of the density matrix
propagator $J_{\rm r}$ from Eq.(\ref{evolred}). In order to solve
the functional integration which defines the reduced propagator,
we perform a saddle point approximation
\begin{equation}
J_{\rm r}[\phi^+_{\rm f},\phi^-_{\rm f},t\vert\phi^+_{\rm i},\phi^-_{\rm i}, t_0]
\approx \exp{ i A[\phi^+_{\rm cl},\phi^-_{\rm cl}]}, \label{prosadle}
\end{equation}
where $\phi^\pm_{\rm cl}$ is the solution of the equation of
motion ${\delta Re A\over\delta\phi^+}\vert_{\phi^+=\phi^-}=0$
with boundary conditions $\phi^\pm_{\rm cl}(t_0)=\phi^\pm_{\rm i}$
and $\phi^\pm_{\rm cl}(t)=\phi^\pm_{\rm f}$.

Even then, it is very difficult to solve this equation
analytically. We are helped by the observation that the ordering
of the field is due to the growth of the long-wavelength unstable
modes. Unstable long-wavelength modes start growing exponentially
as soon as the quench is performed, whereas short-wavelength modes
will oscillate. As a result, the field correlation function
rapidly develops a peak (Bragg peak) at wavenumber $k = {\bar
k}\ll \mu$. Specifically \cite{Karra}, initially as ${\bar k}^2 =
{\cal O}(\mu/\sqrt{t\tau_{\rm q}})$, where $\tau^{-1}_{\rm q}$ is
the quench rate. Assuming that a classical description can be
justified post-hoc, a domain structure forms quickly with a
characteristic domain size $O({\bar k}^{-1})$, determined from the
position of this peak. [As an example, see the numerical results
of \cite{laguna}, where this classical behaviour has been assumed
through the use of stochastic equations (see later)]. With this in
mind, we adopt an approximation in which the system-field contains
only one Fourier mode with $\vec k = \vec k_0 = O({\bar k}^{-1})$,
characteristic of the domain size.

As ``trial'' classical solutions,  we take
\begin{equation}
\phi_{\rm cl}(\vec x, s) =  f(s,t)\cos(k_0 x)\cos(k_0 y)\cos(k_0 z),
\label{classsol}
\end{equation}
where $f(s,t)$ satisfies $f(0,t)= \phi_{\rm i}$ and $f(t,t) =
\phi_{\rm f}$. Eq. (\ref{classsol}) is an exemplary configuration
that mimics domain formation. As intended, it represents domains
of finite size $k_0^{-1}$.  Although such a regular domain
structure is idealised, numerical results (as in \cite{laguna})
suggest that it is a reasonable first step. In practice, after
this detailed motivation, we shall find that the decoherence time
$t_D$ is insensitive to wavelength for $k_0$ small. For the
purpose of calculation, it is sufficient to set $k_0 = 0$, even
though the physical situation requires a finite $k_0$. An exact
matching of $k_0$ to ${\bar k}$ is unnecessary. [We note that, in
\cite{lomplb} and \cite{lomplb2}, we adopted a simpler form
$\phi_{\rm cl}(\vec x, s) =  f(s,t)\cos({\vec k}_0.{\vec x}) $.
Although showing a domain structure in one dimension is less
physical, our conclusions are unchanged although details differ.]

We write $f(s,t)$ as
\begin{equation} f(s,t) = \phi_{\rm i} u_1(s,t) +
\phi_{\rm f} u_2(s,t),\,\, \label{us}\end{equation} where, during
the quench, $u_i(s,t)$ are solutions of the mode equation with
boundary conditions $u_1(0,t) = 1$, $u_1(t,t) = 0$ and $u_2(0,t) =
0$, $u_2(t,t) = 1$. In order to obtain the master equation we must
compute the final time derivative of the propagator $J_{\rm r}$.
After that, all the dependence on the initial field configurations
$\phi^\pm_{\rm i}$ (coming from the classical solutions
$\phi^\pm_{\rm cl}$) must be eliminated. The free propagator,
defined as
\begin{equation}
J_0[\phi^+_{\rm f}, \phi^-_{\rm f}, t\vert \phi^+_{\rm i}, \phi^-_{\rm i}, 0] =
\int_{\phi^+_{\rm i}}^{\phi^+_{\rm f}}{\cal D}\phi^+ \int_{\phi^-_{\rm i}} ^{\phi^-_{\rm f}}
{\cal D}\phi^- \exp\{i [ S_0(\phi^+) -
S_0(\phi^-)]\};
\label{propdeJ0}
\end{equation}
satisfies the general identities \cite{lombmazz,qbm} (which is valid for
instantaneous and for
slow quenches)
\begin{equation}
\phi^\pm_{\rm cl}(s) J_0 =
\Big[ \phi^\pm_{\rm f} [u_2(s,t) - \frac{{\dot u}_2(t,t)}{{\dot
             u}_1(t,t)}u_1(s,t)] \mp  i {u_1(s,t)\over{{\dot u}_1(t,t)}}
\partial_{\phi^\pm_{\rm f}}\Big]J_0.
\label{rel1}
\end{equation}
These identities allow us to remove the initial field
configurations  $\phi^\pm_{\rm i}$, by expressing them in terms of
the final fields  $\phi^\pm_{\rm f}$ and the derivatives
$\partial_{\phi^\pm_{\rm f}}$, and obtain the master equation.

The full equation is very complicated and, as for quantum Brownian
motion, it depends on the system-environment coupling. In what
follows we will compute the diffusion coefficients for the
different couplings described in the previous Section. As we are
solely interested in decoherence, it is sufficient to calculate
the correction to the usual unitary evolution coming from the
noise kernels (imaginary part of the influence action). The result
reads

\begin{equation}
i {\dot \rho}_{\rm r} = \langle \phi^+_{\rm f}\vert
[H,\rho_{\rm r}] \vert \phi^-_{\rm f}\rangle -
i V \sum_j \Gamma_j D_{j}(\omega_0, t) \rho_{\rm r}+ ... \label{master}
\end{equation}
where $j$ labels the different couplings. $D_j$ are the diffusion
coefficients. The ellipsis denotes other terms coming from the
time derivative that do not contribute to the diffusive effects.
$V$ is understood as the minimal volume inside which
there are no coherent superpositions of macroscopically distinguishable
states for the field.  The $\Gamma_j$, which depend on the sums
and differences $\phi^+_{\rm f}\pm\phi^-_{\rm f}$ of the field
amplitudes, have to be calculated case by case.

The effect of the diffusion coefficient on the decoherence process
can be seen considering the following approximate solution to the
master equation
\begin{equation} \rho_{\rm r}[\phi^+, \phi^-; t]\approx
\rho^{\rm u}_{\rm r}[\phi^+, \phi^-; t] ~
\exp \left[-V\sum_j \Gamma_{\rm j} \int_0^t ds ~D_{\rm j}(k_0, s)
\right], \end{equation} where $\rho^{\rm u}_{\rm r}$ is the
solution of the unitary part of the master equation (i.e. without
environment). The system will decohere when the non-diagonal
elements of the reduced density matrix are much smaller than the
diagonal ones.

This decoherence time  $t_{D}$ sets the scale after which we have
a classical system-field configuration, and depends strongly on
the properties of the environment. It is constrained by

\begin{equation}
1 \approx  V \sum_j\Gamma_{\rm j} \int_{0}^{t_{\rm D}} ds ~D_{\rm j}(k_0,s)
\gtrsim V \Gamma_{\rm l} \int_{0}^{t_{\rm D}} ds ~D_{\rm l}(k_0,s), \label{Dsum}
\end{equation}
for any particular $j=l$, and corresponds to the time after which
we are able to distinguish between two different field amplitudes,
inside a given volume $V$. Our conservative choice is that this
volume factor is ${\cal O}(\mu^{-3})$  since $\mu^{-1}$ (the
Compton wavelength) is the smallest scale at which we need to
look.

In the next Section we will compute the corresponding diffusion
coefficients, from which we will estimate the decoherence time
bounds.

\section{Decoherence Times for Slow Quenches} \label{sec:tdec}

Our earliest results from Ref.\cite{lomplb} were for instantaneous
quenches, for which the quench time is effectively $\mu^{-1}\ll
t_{\rm sp}$. {\it A priori}, it was not obvious that this slow
quench preserves these results about the decoherence time, that
$t_D\lesssim t_{\rm sp}$. This is assumed in the Kibble scenario
\cite{Kibble} for domain formation as determined by the appearance
of defects, in which classical defects are assumed to have
appeared as soon as the transition is effected. This was first
addressed by us in a rather schematic way in \cite{lomplb2}, to
which what follows provides a more sophisticated analysis and
extensions to other couplings.

To tackle this problem, we assume that the quench begins at $t=0$
and ends at time $t = 2\tau_{\rm q}$, with $\tau_{\rm q}\gg t_{\rm
r}\sim \mu^{-1}$. At the qualitative level at which we are working
it is sufficient to take $m_{\phi}^2(T_0) = \mu^2$ exactly. Most
simply, we consider a quench linear in time, with temperature
$T(t)$, for which the mass function is of the following form
\cite{bowick}
\[ m^2(t) = m_{\phi}^2(T(t)) = \left\{ \matrix{ \mu^2&\mbox{for} ~
t \le 0\cr \mu^2 -
{t\mu^2\over{\tau_{\rm q}}}& ~~~~~~~~\mbox{for}  ~ 0 < t \le
2\tau_{\rm q}\cr - \mu^2&~\mbox{for} ~ t \ge 2\tau_{\rm q}\cr}
\right.
\]
Note that our $\tau_{\rm q}$ is the inverse quench rate
$T_{\rm c}^{-1}dT/dt|_{T=T_{\rm c}}$, and so differs from that of
\cite{bowick} by a factor of 2.

The field behaves as a free field in an inverted parabolic
potential for an interval of approximately $t_{\rm sp}$ \cite{Karra},
where
\begin{equation}
\langle \phi^2\rangle_{t_{\rm sp}} \sim \eta^2\,\,\, .\label{deftsp2}
\end{equation}
We have rederived $t_{\rm sp}$, using the exact results of
\cite{bowick} (see Appendix).  The end result is that
$t_{\rm sp}$ is the solution to
\begin{equation}
{\eta^2\over C}\simeq {T_{\rm c}\over \mu{\bar t}^2}\exp\bigg[{4\over
3}\bigg({\Delta_0(t_{\rm sp})\over{\bar t}}\bigg)^{3/2}\bigg],
\label{soltsp}
\end{equation}
where $\Delta_0(t) = t - \tau_{\rm q}$,
$C = (64 \sqrt{2}\pi^{3/2})^{-1}$, and ${\bar t}= (\tau_{\rm q}/\mu^2)^{1/3}$
(see details in the Appendix A).

In order to find the master equation (or more
strictly the diffusion coefficient) we follow the same procedure
as in Section III, assuming a dominant wavenumber $k_0$. The classical
solution is given in Eq. (\ref{us}), where $u_i$ satisfy the
mode equation

\begin{equation}
\left[{d^2\over{ds^2}} + k_0^2 + \mu^2 - {\mu^2 s\over{\tau_{\rm q}}}\right]
u_i(s,t) = 0\label{modeeqn}.
\end{equation}
Since we
are neglecting the self-interaction term, our conclusions are only
believable for $t\lesssim t_{\rm sp}$.

The solution of Eq.(\ref{modeeqn}) is given by
              \[
u_1(s,t) =  {-Ai[{\Delta_k (t)\over{{\bar t}}}] Bi[{\Delta_k (s)
\over{{\bar t}}}] + Ai[{\Delta_k (s)\over{{\bar t}}}] Bi[{\Delta_k
(t) \over{{\bar t}}}]\over{Ai[-\omega^2 {\bar t}^2] Bi[{\Delta_k
(t) \over{{\bar t}}}] - Ai[{\Delta_k (t)\over{{\bar t}}}]
Bi[-\omega_0^2 {\bar t}^2]}},
\]

\[u_2(s,t)= {-Ai[{\Delta_k (s)\over{{\bar t}}}] Bi[-\omega_0^2 {\bar t}^2] +
Ai[-\omega_0^2 {\bar t}^2] Bi[{\Delta_k (s) \over{{\bar t}}}]
\over{Ai[-\omega_0^2 {\bar t}^2] Bi[{\Delta_k (t) \over{{\bar
t}}}] - Ai[{\Delta_k (t)\over{{\bar t}}}] Bi[-\omega_0^2 {\bar
t}^2]}} ,
              \] where $Ai[s]$ and $Bi[s]$ are the Airy functions, with
$\Delta_k (s) = s - \omega_0^2{\bar t}^3 $, $\Delta_k (t) = t
- \omega_0^2 {\bar t}^3 $, and $\omega_0^2 = \mu^2 + k_0^2$. We note that
$\Delta_0 (t)= t - \tau_{\rm q}$. In the causal analysis of
Kibble\cite{Kibble} $\bar t$ ($\mu^{-1}\ll \bar
t \ll \tau_{\rm q}$) is the time at which the adiabatic field
correlation length collapses at the speed of light, the earliest
time in which domains could have formed. Our analysis suggests
that this earliest time is not ${\bar t}$, but $t_{\rm sp}$.

\subsection{Quadratic coupling}

We start by considering the biquadratic coupling first, for which the IF is
given by
\[
{\rm Re} \delta A = \frac{g^2}{8} \int d^4 x\int d^4y ~ \Delta_2 (x)
K_{\rm q}(x-y) \Sigma_2 (y), \]
\[
{\rm Im} \delta A = - \frac{g^2}{16} \int d^4x\int d^4y ~ \Delta_2 (x)
N_{\rm q} (x,y) \Delta_2 (y), \]
where $K_{\rm q} (x-y) = {\rm Im} G_{++}^2(x,y) \theta (y^0-x^0)$
is the dissipation kernel and $ N_{\rm q} (x-y) = {\rm Re}
G_{++}^2(x,y)$ is the noise (diffusion) kernel. $G_{++}$ is the
relevant closed-time-path correlator of the $\chi$-field at
temperature $T_0$. We have defined
$$\Delta_2 ={1\over{2}}(\phi^{+2} - \phi^{-2}) ~~~;~~~ \Sigma_2 ={1\over{2}}
(\phi^{+2} + \phi^{-2}).$$

We can formally find the master equation in the same way that we
did in Section III. It is given by Eq.(\ref{master}), with a
diffusion coefficient of the form (details on how to get this
coefficient can be found in Appendix B)
\begin{equation} D_{\rm
qu}(k_0,t) = \int_0^t ~ ds ~ u(s,t)~ F(k_0,s,t),\label{dqsq}
\end{equation} with
$$u(s,t) =\bigg[u_2(s,t) - \frac{{\dot u}_2(t,t)}{{\dot
              u}_1(t,t)}u_1(s,t)\bigg]^{2},$$ and
$$
F(k_0,s,t) = {1\over{64}}\left\{{\rm Re}G_{++}^2(0; t-s) +
{3\over{2}}{\rm Re} G_{++}^2(2k_0; t-s) +
{3\over{4}}{\rm Re}G_{++}^2(2\sqrt{2}k_0; t-s) + {1\over{8}}{\rm Re}G_{++}^2(2\sqrt{3}k_0; t-s)
\right\}.$$
It is only in $u(s,t)$ that the slow quench is apparent. $G^2_{++}(k, t-s)$ is the
Fourier transform of the square of the Feynman propagator ($\chi$ propagator).

In the high temperature limit ($T \gg \mu$), the explicit
expression for the kernels can be shown, with a
little labour, to be
\begin{equation}
{\rm Re}G_{++}^2(k; t-s)={T_{\rm c}^2\over{64\pi^2}}
{1\over{k}} \int_0^\infty dp~{p\over{p^2 + \mu^2}}
\int_{\sqrt{\vert p - k\vert^2 + \mu^2}}^{\sqrt
{\vert p + k\vert^2 + \mu^2}}{du\over{u}}
~\cos[(\sqrt{p^2 + \mu^2}+ u) (t-s)], \label{D1} \end{equation}
and
\begin{equation}
{\rm Re}G_{++}^2(0; t-s)={T_{\rm c}^2\over{64\pi^2}} ~
\int_0^\infty dp~{p^2\over{(p^2 + \mu^2)^2}} \cos[2\sqrt{p^2 +
\mu^2}(t-s)], \label{D} \end{equation}
where $\mu$ is the thermal $\chi$-field mass at temperature $T\sim
T_{\rm c}$. In our scheme, this is approximately the cold $\chi$
mass.

It is because the $\chi$-field propagator is unaffected by the
$\phi$-field interactions that the detail of the expressions
(\ref{D1})-(\ref{D}) are possible. Fortunately, it is inessential. We see
that, for times $\mu t\gtrsim 1$, the behaviour of $D_{\rm
qu}(k_0, t)$ is dominated by the exponential growth of $u(s,t)$,
and the integral in Eq.(\ref{dqsq}) by the interval $s\approx 0$. Indeed,
it is easy to prove that $u(0,t) = ({{\dot u}_2(t,t)\over{{\dot
u}_1(t,t)}})^2 \gg 1$, while $u(t,t) = 1$. Therefore, although the
expessions in Eqs.(\ref{dqsq}), (\ref{D1}), and (\ref{D}) are complicated,
we can approximate the whole diffusion coefficient by

\begin{equation}D_{\rm
qu}(k_0, t) \approx F(k_0, 0,t) ~ u(0,t) ~ \int_0^{\omega_0^{-1}}
ds ~ {u(s,t)\over{u(0,t)}} \label{D0}
\end{equation}
where we used the fact that $F(k_0, s,t)$ is bounded at $s=0$.

We will assume large $\Delta_k (t)$ (and $\Delta_k (s)$), which
means $\Delta_k (t), \Delta_k (s) \gg {\bar t}$. This condition is
satisfied provided $s$ is larger and not to close to $\omega_0^2
\tau_{\rm q}/\mu^2$, and allow us to use the asymptotic
expansions of the Airy functions and their derivatives for the
evaluation of $u_i(s,t)$. This will be justified posthoc. In particular
we obtain
\begin{eqnarray}
u(0,t)&\simeq&{1\over 4 \omega_0 {\bar t}}\sqrt{\Delta_k (t)\over
{\bar t}}
\exp{\left\{ {4\over{3}}\left({\Delta_k (t)\over{{\bar t}}}\right)^{3\over{2}}
\right\}}\nonumber\\
\dot u(0,t)&\simeq&{1\over 2 \omega_0^2{\bar t}^3}\sqrt{\Delta_k
(t) \over {\bar t}} \exp{\left\{ {4\over{3}}\left({\Delta_k
(t)\over{{\bar t}}}\right)^{3\over{2}} \right\}}\label{uupunto}.
\end{eqnarray}
Therefore,  it
is straightforward  to check that in these cases the integral in
Eq.(\ref{D0}) is given by

\[
\omega_0 ~\int_0^{\omega_0^{-1}}ds \left[1 + s {{\dot u}(0,t)
\over{u(0,t)}} + ...~ \right] \approx \omega_0
~\int_0^{\omega_0^{-1}}ds \left[ 1 + s {2\over{\omega_0 {\bar
t}^2}} + ... \right] ,
\]
and it is ${\cal O}(1)$, due to the fact that $\mu {\bar t} \gg
1$.

We can estimate the decoherence time for the quadratic coupling as
\begin{equation}
1 \gtrsim V\Gamma\int_0^{t_{D}} dt ~ D_{\rm qu}(k_0,t),
\end{equation} where $V$ is the decoherence volume, as before.

In terms of the dimensionless fields $\bar\phi = (\phi^+ +
\phi^-)/2\mu,$ and $ \delta = (\phi^+ - \phi^-)/2\mu$ it follows
that $\Gamma_{\rm qu} = g^2\mu^4 \bar\phi^2 \delta^2$. In
order to quantify the decoherence time we have to fix the values
of $\delta$, and $\bar\phi$.  Inside the volume $V$ we do not
discriminate between field amplitudes which differ by $ {\cal
O}(\mu) $, and therefore, for the sake of argument, we take
$\delta = 1$. For $\bar\phi$ we set $\bar\phi^2 =\alpha /\lambda$,
where $\lambda\leq\alpha\leq 1$ is to be determined
self-consistently. It is necessary to be as precise as this in the
first instance, since $t_{\rm sp}$ and $t_D$ only differ by
logarithms. It is important that the prefactors in the arguments
of these logarithms are determined carefully, given that terms
nominally ${\cal O}(1)$ can be small. Once we have established
that the difference is large enough, this
artificial precision can be dropped. Posthoc it is sufficient to
take $\delta \sim {\cal O}(1)$ and $\bar\phi^2\sim {\cal O}(\alpha
/\lambda)$, as we did in \cite{lomplb} and \cite{lomplb2}.

Then, for quadratic interactions alone the decoherence time reads
\begin{equation}
\exp{\left\{ {4\over{3}}\left({\Delta_k (t_D)\over{\bar t
}}\right)^{3\over{2}} \right\}} \approx
4 . 10^4{\omega_0^2\over{\lambda T_{\rm c}^2 \alpha}}.
\end{equation}

Using $\langle\phi^2\rangle$ of
(\ref{phi2}) we obtain $\alpha^2 \approx 7 (\mu/T_{\rm c})(\mu \tau_{\rm
q})^{-2/3}$. Thus the decoherence time satisfies
\begin{equation}
\exp{\left\{ {4\over{3}}\left({\Delta_k (t_D)\over{\bar t
}}\right)^{3\over{2}} \right\}} \approx
2 . 10^3 {\omega_0^2\over{\mu^2}}{\eta^2\over{T^{{3\over{2}}}_{\rm
c}}} {(\mu\tau_{\rm q})^{{1\over{3}}}\over{\mu^{1\over{2}}}},
\end{equation}
for long-wavelength modes.

If we compare it with the spinodal time of (\ref{soltsp}), we get,
\begin{equation}
(\mu \Delta_0 (t_{\rm sp}))^{{3\over{2}}} - (\mu \Delta_k
(t_{D}))^{{3\over{2}}} \approx {3\over{4}}~ \sqrt{\mu \tau_{\rm
q}}~ \left\{2 \ln {\mu\over{\omega_0}} +
\ln \left[{1\over{2}}\left({T_{\rm
c}\over{\mu}}\right)^{1\over{2}} (\mu\tau_{\rm
q})^{1\over{3}}\right]\right\} > 0. \label{ts2}
\end{equation}
We see that the numerical prefactor in the argument of the
logarithm is, indeed, ${\cal O}(1)$, and our concern for precision
was unnecessary, in retrospect. Our approximation scheme
depends, as for the instantaneous
quench, on the peaking of the power in the field fluctuations at
long-wavelength $k_0\ll\mu$ by time $t_{\rm sp}$, and it is
sufficient to take $k_0\approx 0$. In this case (\ref{ts2})
becomes
\begin{equation}
(\mu \Delta_0(t_{\rm sp}))^{{3\over{2}}} - (\mu \Delta_0
(t_{D}))^{{3\over{2}}} \approx {3\over{4}}~ \sqrt{\mu \tau_{\rm q}}~
\ln \left[{1\over{2}} \left({T_{\rm
c}\over{\mu}}\right)^{1\over{2}} (\mu\tau_{\rm
q})^{1\over{3}}\right] > 0, \label{ts3}
\end{equation}
from which $t_D < t_{\rm sp}$ follows, as in the instantaneous
case. The inclusion of further interactions, including the
self-interaction with short-wavelength ($k > \mu$) modes can only
reduce $t_D$ further.

We should point out that, whereas peaking is inevitable for the
instantaneous quench for weak enough coupling, it is not the case
for very slow quenches. In such cases our approximations break
down and a different analysis is required. The details are rather
messy, but a sufficient condition for our approximation to be
valid is that $\mu\tau_{\rm q} \lesssim\eta/\mu$\cite{Karra}.
Tighter, but less transparent bounds can be given\cite{Karra}.

When these bounds are satisfied the minimum wavelength for which
the modes decohere by time $t_{\rm sp}$ can be shown\cite{lomplb2}
to be shorter than that which characterises domain size at that
time. Although we can talk loosely, but sensibly, about a
classical domain structure at time $t_{\rm sp}$ we cannot yet talk
about classical defects on their boundaries, as the naive picture
might suggest. Defects (in this case, walls) are described by
shorter wavelength modes ($k\lesssim\mu$). Nonetheless, the
classical domain structure is sufficient to determine their
density \cite{lomplb2}.

\subsection{Bilinear and linear couplings}

Another possibility that needs to be considered is that of a
bilinear coupling of the $\phi$-field to the environment. This
interaction preserves the $\phi\rightarrow -\phi$ symmetry of $
S_{\rm syst}[\phi ]$ in which, for simplicity we continue to take
bilinear couplings equal, as $g_{\rm a}'=g'/\sqrt{N}$. We treat
this interactions as an {\it additional} set of interactions to
the biquadratic interactions, whereby $T_{\rm c}$ is qualitatively
unchanged.

The influence action is still obtained from Eq.(\ref{inflac}), as

\begin{equation}
{\rm Re} \delta A_{\rm bilin} =  {g'^{2}\mu^2\over{8}} \int d^4 x\int d^4y ~
\Delta_2 (x) K_{\rm b}(x-y) \Sigma_2 (y), \end{equation}
and
\begin{equation}
{\rm Im} \delta A_{\rm bilin} = -  {g'^{2}\mu^2\over{16}} \int d^4x\int d^4y ~
\Delta_2 (x) N_{\rm b}(x-y) \Delta_2 (y),\end{equation} where

\begin{equation}K_{\rm b} = {\rm Re}G_{++}(x,y)\theta (y^{0}-x^{0})
,\,\,\,\,\,\,N_{\rm b} = {\rm Im}G_{++}(x,y).\end{equation}

The temporal diffusion
coefficient is now given by
\begin{equation}
D_{\rm bilin}(k_0, t) = \int_0^t ds ~ u(s,t) ~ F_{\rm bilin}(k_0, s, t), \label{diff2}
\end{equation}
where
$$
 F_{\rm bilin}(k_0, s, t) = {1\over{64}}\left\{{\rm Im}G_{++}(0; t-s) +
{3\over{2}}{\rm Im} G_{++}(2k_0; t-s) +
{3\over{4}}{\rm Im}G_{++}(2\sqrt{2}k_0; t-s) + {1\over{8}}{\rm Im}G_{++}(2\sqrt{3}k_0; t-s)
\right\}.$$
with
\begin{equation}
{\rm Im} G_{++}(k; s) = \frac{T_{\rm c}}{4}\frac{\cos[\sqrt{k^{2}
+ \mu^2 }s]}{k^{2} + \mu^2}.
\end{equation}
Following the same arguments as in the previous section, we can
evaluate the diffusion coefficient by

\begin{equation}D_{\rm bilin}(k_0, t) \approx F_{\rm bilin}(k_0, 0, t){u(0, t) \over{\omega_0}},
\end{equation}
where $u(0,t)$ is given in Eq.(\ref{uupunto}). Thus,

\begin{equation}D_{\rm bilin}(k_0, t) \approx 10^{-2}
\frac{T_{\rm c}}{\mu^2\omega_0} ~u(0, t).
\end{equation}

For the unstable long-wavelengths ($k_0^2 < \mu^2/2$
approximately) we
find, for times $\mu t\gtrsim 1$, that $D_{\rm
bilin}(k_0, t)$ again shows the exponential growth
\begin{equation}
D_{\rm bilin}(k_0, t) \approx \frac{3 . 10^{-3} T_{\rm c}}{{\bar t}\mu^2
\omega_0^2} \sqrt{{\Delta_k (t)\over{{\bar t}}}} \exp\left\{
{4\over{3}}\left( {\Delta_k (t)\over{{\bar t}}}\right)^{3\over{2}}
\right\}. \label{Dyu2}
\end{equation}
It follows that $\Gamma_{\rm bilin} = g'^2\mu^4 \bar\phi^2
\delta^2$. As a result, the contribution to the decoherence time from bilinear
interaction (for long-wavelength
modes) can be evaluated as
\begin{equation}
\exp\left\{ {4\over{3}}\left( {\Delta_k (t_D)\over{{\bar t}}}
\right)^{3\over{2}} \right\} \approx 6 . 10^2 {\omega_0^2\over{\lambda T_{\rm
c}\mu \alpha}}.
 \end{equation}

The value of
$\alpha$ is again determined from the condition that, at time
$t_D$, $\langle \phi^2\rangle\sim\alpha\eta^2$. That is, $\alpha^2 =
0.1 (\mu \tau_{\rm q})^{-2/3}$. Thus, decoherence time is given by
\begin{equation}
 \exp\left\{ {4\over{3}}\left( {\Delta_k
(t_D)\over{{\bar t}}}\right)^{3\over{2}} \right\} \approx
3 . 10^2 {\omega_0^2\over{\mu^2}} {\eta^2\over{\mu T_{\rm
c}}}(\mu \tau_{\rm q})^{1\over{3}}.\end{equation}

From (\ref{soltsp}),
$t_D$ and $t_{\rm sp}$ are related by

\begin{equation}
(\mu\Delta_0 (t_{\rm sp}))^{{3\over{2}}} - (\mu\Delta_0
(t_{D}))^{{3\over{2}}} \approx {3\over{4}}~ \sqrt{\mu \tau_{\rm
q}}~ \ln \left[3 (\mu \tau_{\rm
q})^{1\over{3}}\right],
\end{equation}
which is positive for a sufficiently slow quench.

As we have already observed, early studies of decoherence were
confined largely to quantum mechanical systems, for which the
environment was typically a collection of harmonic oscillators, to
which the system coupled linearly. Such systems have the virtue of
exact solvability (or closed equations) and have been very
instructive. However, in the context of quantum field theory
linear terms are usually a signal of an inappropriate choice of
field basis. Further, as we are considering models with
spontaneous symmetry breaking, linear couplings with external
fields are not a natural choice since they break the vacuum
degeneracy. We include isolated linear couplings for completeness.
Again, choosing couplings equal,  as $g''_{\rm a}=g''/\sqrt{N}$,
and defining $\Delta_{\rm 1} = (\phi^{+} - \phi^{-})/2$ and
$\Sigma_{\rm 1}= (\phi^{+} + \phi^{-})/2$, we are able to write
the real and imaginary parts of the influence functional as

\begin{equation}
{\rm Re} \delta A_{\rm lin} =  {g''^{2}\mu^4\over{8}} \int d^4 x\int d^4y ~
\Delta_{\rm 1} (x) K_{\rm b}(x-y) \Sigma_{\rm 1} (y), \end{equation}
and
\begin{equation}
{\rm Im} \delta A_{\rm lin} = -  {g''^{2}\mu^4\over{16}} \int d^4x\int d^4y ~
\Delta_{\rm 1}(x) N_{\rm b}(x-y) \Delta_{\rm 1}(y).\end{equation}

For times $\mu t\gtrsim 1$ the diffusion contribution to the master equation is

\begin{equation}
D_{\rm lin}(k_0, t) \approx \frac{T_{\rm c}}{8 \omega_0^3}~\sqrt{u(0, t)}.
\label{DL}
\end{equation}

Note that the exponent is only {\it half} that of the quadratic and
bilinear interactions.

In this case, it can be shown that the decoherence time comes from

\begin{equation}1 \approx {{V \Gamma_{\rm lin} T_{\rm c}}\over{16 \omega_0^3}}
\sqrt{{{\bar t}\over{\omega_0}}} \left[- \Gamma [{5\over{6}}] +
\Gamma [{5\over{6}}, -{2\over{3}}({\Delta_k(t_D)\over{{\bar
t}}})^{3\over{2}}]\right], \label{incomplete}\end{equation} where
$ \Gamma [a,z]$ is the incomplete Gamma function and, in this case
$\Gamma_{\rm lin} = {1\over{4}} g''^2 \mu^2 \delta^2$. As $t_D \gg {\bar t}$, we can
approximate Eq.(\ref{incomplete}) and obtain,

\begin{equation} \exp\left\{ {2\over{3}}\left(
{\Delta_k(t_D)\over{{\bar t}}}\right)^{3\over{2}} \right\} \approx
{16\over{9}}{\omega_0^3\over{\mu^3}}~ {\eta^4\over{\mu^4}}
~{1\over{T_{\rm
c}}}{\omega_0^{1\over{2}}\mu^{1\over{3}}\over{\tau_{\rm
q}^{1\over{6}}}}
.\end{equation}

The decoherence time associated to the linear interaction term, can be written as

\begin{equation}\Delta_k(t_D)^{3\over{2}} \approx {3\over{2}} {\bar t}^{3\over{2}} ~
\left\{3\ln {\omega_0\over{\mu}} + \ln
{16\over{9}} {\eta^4\over{\mu^4}}
~{1\over{T_{\rm
c}}}{\omega_0^{1\over{2}}\mu^{1\over{3}}\over{\tau_{\rm
q}^{1\over{6}}}}\right\}.
\end{equation}

$t_D$ and $t_{\rm sp}$ are related by

\begin{equation}
(\mu\Delta_0(t_{\rm sp}))^{{3\over{2}}} - (\mu\Delta_0(t_{D}))^{{3\over{2}}}
\sim {3\over{4}}~ \sqrt{\mu \tau_{\rm q}}~~ \ln \left[10^3 {\mu^2\over{\eta^2}}
(\mu\tau_{\rm q})^{5\over{6}}\right].
\end{equation}

In this case, due to the bound on the quench time ($\tau_{\rm
q}\leq \eta /\mu^2$), we find that it looks as if $t_{D} > t_{\rm
sp}$. However this result is not believable as it stands, since
the diffusion coefficient has been computed assuming $t\lesssim
t_{\rm sp}$. Whatever, the rapid decoherence of the biquadratic
and other couplings is not present. This shows how adopting linear
coupling to an environment in mimicry of quantum mechanics can be
misleading.

\subsection{Comparison between different couplings} \label{sec:weak}

In our present model the environment fields $\chi_{\rm a}$ are not the
only decohering agents. The environment is also constituted by the short-wavelength modes
of the self-interacting field $\phi$.
Therefore, we split the field as $\phi = \phi_< + \phi_>$, and define the system ($\phi_<$)
by those modes with wavelengths longer than the critical value $\Lambda^{-1}$, while the
bath or environment-field ($\phi_>$) contains wavelengths shorter than $\Lambda^{-1}$. In
order to consider only the unstable modes inside our system, we will set this critical
scale $\Lambda$ of the order of $\mu$. In practice, where the separation is made exactly is
immaterial\cite{Karra} by time $t_D$, when the power of the
$\phi$-field fluctuations is peaked at $k_0\ll\mu$. The effect is
to give a separation of system from environment through the
decomposition of $S[\phi , \chi ]$ of (\ref{action0}) introducing
an interaction terms of the form $\phi_< \phi_>^3$, $\phi_<^2 \phi_>^2$, and $\phi_<^3 \phi_>$.
The relevant interaction term is
\begin{equation}
S_{\rm couple}[\phi_{<} ,\chi ] = S_{\rm int}[\phi_{<} ,\chi ]-{\lambda\over{4}}\int
d^4x \,(\phi_{<}(x) \phi_{>} (x))^2.
\label{action}\end{equation}
All terms omitted in the expansion \cite{lombmazz,greiner} do not contribute to the
one-loop calculations for the long-wavelength modes that we shall now consider.
\footnote{Strictly speaking, for $\mu/3<k_0<\mu$ one should include an additional term
proportional to $\phi_<^3\phi_>$ in the interaction Lagrangian.
See Ref. \cite{lombmazz} for details.}

The net consequence of the separation of the long-wavelength system modes $\phi_<$ of $\phi$
from the short-wavelength modes $\phi_>$ of the environment through the interaction action
$S_{\rm couple}$ of (\ref{action}) gives an additional one-loop contribution $D_{\phi}(k_0,t)$
 to the diffusion function with the same $u(s,t)$ and the same form as in (\ref{dqsq}).
However, $G_{\pm\pm}$ is now constructed from the short-wavelength
modes $\phi_>$ of the $\phi$-field as it evolves from the top of
the potential hill. Without the additional powers of $N^{-1}$ to
order contributions the one-loop calculation is unreliable. In
fact, we would not expect the inclusion of the $\phi$-field to
give a qualitative change at one-loop level. In the first
instance, the approximation (\ref{D0}) remains valid, and the
1-loop term is driven by the exponential growth of $u(0,t)$.
Secondly, the effect is that the short-wavelength modes have been
kept at the initial temperature $T_0$, on the grounds that passing
through the transition quickly has no effect on them. That is,
with $g\simeq\lambda$ and no $1/N$ factor, the short-wavelength
$\phi$ modes give a contribution comparable, qualitatively, to
{\it all} the explicit environmental fields put together. At an
order of magnitude level there is no change, since the effect is
to replace $g^2$ by $g^2 + O(\lambda^2) = O(g^2)$.

However, since the contribution of $D_{\phi}(k_0,t)$ to the
overall diffusion function is positive we can derive an {\it
upper} bound on the decoherence time $t_D$ from the reliable
diffusion functions $D_{\rm qu}(k_0,t)$, $D_{\rm bilin}(k_0,t)$
(and $D_{\rm lin}(k_0,t)$). Let us suppose we have a theory where
the three considered couplings with the environment are present.
Eq.(\ref{Dsum}) will be satisfied because one of its terms will
grow faster than the others, rather than because many terms will
each give a small fraction of unity. Specifically, we have (up to
numerical factors $O(1)$)
\begin{eqnarray}
&& \Gamma_{\rm qu} \int_{0}^{t_{D}} ds ~D_{\rm qu}(k_0,s): \Gamma_{\rm bilin} \int_{0}^{t_{D}}
ds ~D_{\rm bilin}(k_0,s): \Gamma_{\rm lin} \int_{0}^{t_{D}} ds ~D_{\rm lin}(k_0,s) \nonumber
\\
&\sim& g\,\,:\,\,
g'\bigg(\frac{\mu}{T_{\rm c}}\bigg)\,\,:\,\, g''\frac{1}{\delta^2}\bigg(\frac{\mu}
{T_{\rm c}}\bigg)\,\exp\left\{{2\over{3}}\left({\Delta_0(t_D)
\over{\bar t}}\right)^{3\over{2}}\right\}.
\end{eqnarray}
Since $T_{\rm c}\gg\mu$ and $\delta \approx O(1)$ then, for $\mu t_D\gg 1$, we have
\begin{equation}
\Gamma_{\rm qu} \int_{0}^{t_{D}} ds ~D_{\rm qu}(k_0,s)\,\, \gg\,\, \Gamma_{\rm bilin}
\int_{0}^{t_{D}} ds ~D_{\rm bilin}(k_0,s)\,\,\gg\,\, \Gamma_{\rm lin} \int_{0}^{t_{D}} ds ~
 D_{\rm lin}(k_0,s),
\end{equation}
if $g,g'$ and $g''$ are comparable. Thus,  it is sufficient to
evaluate the constraint on the time $t_{D}$ from this biquadratic
interaction alone.

\section{Final remarks}
\label{sec:final}

We first summarize the results contained in this paper, in which we have shown how the
environment leads to the decoherence of the order parameter after a transition.
After the integration over the scalar environment-fields $\chi_{\rm a}$ in
Section II, we have obtained the coarse-grained effective
action (CGEA) for the system (field $\phi$). From the imaginary part of the
CGEA we obtained the diffusion coefficient of the master equation, at 1-loop and in
the high temperature environment limit. Terms omitted are relatively $O(N^{-1/2})$
for $N$ weakly coupled environmental fields. Subsequently, we evaluated the
decoherence time for the long-wavelength modes of the system-field
(for each of the different couplings with the environment considered) for slow quenches. This
decoherence time depends on the coupling between system and bath, the self-coupling
of the system (through the environment temperature $T$), and the mass. In our model,
we have shown that the decoherence time is in general  {\it smaller} than the spinodal time.

We stress that the inequality $t_D < t_{\rm sp}$ is insensitive to the
strength of the couplings, for weak coupling. The reason is twofold.
Firstly, there is the effect that $\Gamma\propto
T_0^2$, and $T_0^2 \propto\lambda^{-1}$ is non-perturbatively large for a
phase transition.  Secondly,
because of the non-linear coupling to the environment, obligatory for
quantum field theory,
$\Gamma\propto{\bar\phi}^2$. The completion of the transition finds
${\bar\phi}^2\simeq\eta^2\propto\lambda^{-1}$ also non-perturbatively
large. This suggests
that $\Gamma$, and hence $t_D$, can be independent of $\lambda$.  In fact,
the situation is a little more complicated, but the inequality holds.

This arises because the diffusion coefficients, which trigger
classical behaviour when they become large enough, are controlled
by the exponential growth of the unstable modes. It is this same
exponential growth that determines $t_{\rm sp}$. This result
provides a post-hoc justification of our assertion
\cite{Karra,Rivers1} that the spinodal time sets the scale for the
onset of classical behaviour (in open systems).

Our emphasis has been on the many weak environments because of the
control that this gives us on establishing a robust upper bound on
$t_D$. However, we noted earlier that their total contribution was
qualitatively that of the short-wavelength modes of the $\phi$
field alone. Environmental fields are an important feature of the
early universe, but even had we not included them we would have
expected a similar result from the one-loop couplings of 
short-wavelength to long-wavelength $\phi$ modes. Although, in this
case, we have no way to control the higher loop terms, it is quite
probable that our prescription of the environments is
unnecessarily detailed, and early decoherence is a general
feature.

Ideally, we could extend our ideas to gauge theories. This is
beyond the scope of this paper, but we would like to emphasize an
important point. As diffusion is additive (each extra term in the
interaction action gives additional diffusion coefficients to the
master equation), the inclusion of further couplings and fields
can only reduce $t_D$. Further, the coupling of short to 
long-wavelength modes of the global theory is omnipresent. However,
theories with derivative couplings (scalar QED, for example) tend
to produce small additional diffusive effects at low $\vec k$
(scaled by $k$). It is for this reason that, if scalar fields are
present, there is no need include gauge field interactions in
order to estimate an upper bound to the decoherence time.

The results obtained in this paper also justify in part 
the use of phenomenological
stochastic equations to describe the dynamical evolution of the
system field, as we will now discuss. As it is well known
\cite{lombmazz,greiner}, for $\phi^2\chi^m$ interactions one can
regard the imaginary part of $\delta A$ as coming from a noise
source $\xi_{\rm m}(x)$, with a Gaussian functional probability
distribution. Taking the biquadratic coupling to the external
environment $\chi$-field first leads to a noise, termed $\xi_2$,
say, with distribution
\begin{equation}
P[\xi_{2}(x)]= N_{\xi_{2}} \exp\bigg\{-{1\over{2}}\int d^4x\int d^4y ~\xi_{2} (x)
\Big[ g^2 N_{\rm q}\Big]^{-1}\xi_{2} (y)\bigg\}, \end{equation}
where $N_{\xi_{2}}$ is a normalization factor.  This enables us to write the
imaginary part of the influence action as a functional integral over the Gaussian
field $\xi_{2} (x)$ %
\begin{eqnarray}
\int {\cal D}\xi_{2} (x) P[\xi_{2} ]&&\exp{\left[ -i \bigg\{\Delta_{2} (x) \xi_{2} (x)
\bigg\}\right]}\nonumber \\ &=& \exp{\bigg\{-i\int d^4x\int d^4y \ \Big[\Delta_{2}(x)
~g^2 N_{\rm q}(x,y)~ \Delta_{2}(y)\Big] \bigg\}}.\end{eqnarray} %
In consequence, the CGEA can be rewritten as %
\begin{equation}A[\phi^+,\phi^-]=-{1\over{i}} \ln  \int {\cal D} \xi_{2}
  P[\xi_{2}]
\exp\bigg\{i S_{\rm eff}[\phi^+,\phi^-, \xi_{2}]\bigg\}, \end{equation}
where
\begin{equation}
S_{\rm eff}[\phi^+,\phi^-,\xi_{2}]= {\rm Re} A[\phi^+,\phi^-]- \int d^4x\Big[\Delta_{2}
(x) \xi_{2}(x) \Big].
\end{equation}

Therefore, taking the functional variation %
\begin{equation}
\left.{\delta S_{\rm eff}[\phi^+,\phi^-, \xi_{2}]\over{\delta
\phi^+}}\right\vert_{\phi^+=\phi^-}=0, \label{statphase}
\end{equation}
we are able to obtain the ``semiclassical-Langevin'' equation for the system-field
\cite{lombmazz,greiner}

\begin{eqnarray}
\Box \phi (x) - {\tilde\mu}^2 \phi +
{{\tilde\lambda}\over{6}}\phi^3(x) &+& g^2 \phi (x) \int d^4y ~
K_{\rm q}(x-y)~ \phi^2(y) = \phi (x) \xi_{2}(x)\label{lange2}
,\end{eqnarray} where  ${\tilde\mu}$ and $ {\tilde\lambda}$ are
constants ``renormalised'' because of the coupling with the
environment.  Since (\ref{lange2}) is, from (\ref{statphase}), the
stationary phase approximation, it is only valid once the system
has become classical. It can be used to establish domain formation
only because $t_D < t_{\rm sp}$.

Each part of the environment that we include leads to a further 'dissipative' term on the
left hand side of (\ref{lange2}) with a countervailing noise term on the right hand side.
Once we include the interactions between $\phi_<$ and $\phi_>$ at one-loop level we get a similar
equation to (\ref{lange2}) from the term $S_{\rm couple}$ alone (see Eq.(\ref{action})).
However, although the $\phi_<\phi^3_>$ and $\phi^3_<\phi_>$ terms were ignorable in the
bounding of $t_D$, in the Langevin equations they give further terms, with quadratic
$\phi^2\xi_3$ noise and linear noise $\xi_1$ respectively.

The inclusion of bilinear interactions leads to the inclusion of
further terms in (\ref{lange2}) of the same form. For the linear
interaction with the environment (to the exclusion of
self-interaction) we do recover the {\it additive} noise that has
been the basis for stochastic equations in relativistic field
theory that confirm the scaling behaviour of Kibble's and Zurek's
analysis \cite{laguna}.

For times later than $t_{\rm sp}$, neither perturbation theory nor
more general non-Gaussian methods are valid. It is difficult to
imagine an {\it ab initio} derivation of the dissipative and noise
terms  from the full quantum field theory. In this sense, a
reasonable alternative is to analyze   phenomenological stochastic
equations numerically and check the robustness of the predictions
against different choices of the dissipative kernels and of the
type of noise. We stress that this is only possible because $t_D <
t_{\rm sp}$.  This will be considered further  elsewhere
\cite{nunopedroray}.

Finally, we see that the role initially attributed by Kibble \cite{Kibble,kzm} (and subsequently
by others e.g.\cite{joao}) to the Ginzburg regime is just not present.

\acknowledgments F.C.L. and F.D.M. were supported by Universidad de Buenos Aires, CONICET,
Fundaci\'on Antorchas and ANPCyT. R.J.R.
was supported, in part, by the COSLAB programme of the European Science Foundation.

\appendix
\section{The spinodal time}
\label{rw-appendix}

In this Section, we show the estimation of the spinodal time $t_{\rm sp}$
defined from $\langle \phi^2 \rangle_{t=t_{\rm sp}} \sim \eta^2$.

The equation of motion for the mode ${\cal U}_k(t)$, with
wavenumber $k$ is, in the quench period,
\begin{equation}
 \left[{d^2\over{ds^2}} + k^2 + \mu^2 - {\mu^2 s\over{\tau_{\rm q}}}\right]
{\cal U}_k(t) = 0, \label{airy1}
 \end{equation}
subject to the boundary condition ${\cal U}_k(t) = e^{-i\omega t}$
for $t\leq 0$, where $\omega^2 = \mu^2 + k^2$.

Instead of the simple exponentials of the instantaneous quench,
${\cal U}_k(t)$ has solution
\begin{equation}
{\cal U}_k(t) =  a_k Ai[{\Delta_k (t)\over{{\bar t}}}] + b_k
Bi[{\Delta_k (t) \over{{\bar t}}}], \label{airy1sol}
\end{equation}
with $Ai[s]$, $Bi[s]$ the Airy functions; $\Delta_k (t) = t -
\omega^2 {\bar t}^3 $ and ${\bar t} = (\tau_{\rm q}/\mu^2)^{1/3}$.
Note that $\Delta_0 (t) = t - \tau_Q$, the time since the onset of
the transition. In the causal analysis of Kibble\cite{Kibble}
${\bar t}$ ($\mu^{-1}\ll {\bar t}\ll \tau_{\rm q}$) is the time at
which the adiabatic field correlation length collapses at the
speed of light, the earliest time in which domains could have
formed. Our analysis suggests that this earliest time is not
${\bar t}$, but $t_{\rm sp}$.

It is straightforward to establish a relationship between ${\bar
t}$ and $t_{\rm sp} > {\bar t}$. The constants of integration in
(\ref{airy1sol}) are
\begin{eqnarray}
a_k &=& \pi [Bi'(-\omega^2{\bar t}^2) + i\omega{\bar t}Bi
(-\omega^2{\bar t}^2)],\\
\nonumber
 b_k &=& - \pi [Ai'(-\omega^2{\bar t}^2) +
i\omega{\bar t}Ai (-\omega^2{\bar t}^2)].
\end{eqnarray}
It follows that, when $\Delta_k (t)/{\bar t}$ is large, then
\begin{eqnarray}
|{\cal U}_k(t)|^2&\approx& \omega{\bar
t} \bigg({{\bar t}\over\Delta_k(t)} \bigg)^{1/2}\exp\bigg[{4\over
3}\bigg({\Delta_k(t)\over{\bar t}}\bigg)^{3/2}\bigg] \nonumber
\\
&\approx& \mu{\bar t} \bigg({{\bar
t}\over\Delta_0(t)} \bigg)^{1/2}\exp\bigg[{4\over
3}\bigg({\Delta_0(t)\over{\bar t}}\bigg)^{3/2}\bigg]e^{-k^2/{\bar
k}^2},
\end{eqnarray}
where ${\bar k}^2 = {\bar t}^{-3/2}(\Delta_0(t))^{-1/2}/2$.

For large initial temperature $T_0 = O(T_c)$, we find the power
spectrum for field fluctuations peaked around ${\bar k}$, and
\begin{equation}
\langle \phi^2\rangle_{t}\approx {T_0\over 2\pi^2\mu^2}\int
k^2\,dk\,|{\cal U}_k(t)|^2\approx {CT_0\over\mu{\bar
t}^2}\bigg({\Delta_0(t)\over{\bar t}}
\bigg)^{-5/4}\exp\bigg[{4\over 3}\bigg({\Delta_0(t)\over{\bar
t}}\bigg)^{3/2}\bigg]. \label{phi2}
\end{equation}
We have intentionally included the prefactor $C$ to show that
terms, nominally $O(1)$, can in fact be large or small (in this
case $C= (64\sqrt{2}\pi^{3/2})^{-1}=O(10^{-3}))$. Note that,
although the unstable modes have a limited range of $k$-values,
increasing in time, this is effectively no restriction when
$\Delta_0(t)/{\bar t}$ is significantly larger than unity.

Finally, we obtain

\begin{equation}
{\eta^2\over C'}\simeq {T_{\rm c}\over \mu{\bar t}^2}\exp\bigg[{4\over
3}\bigg({\Delta_0(t_{\rm sp})\over{\bar t}}\bigg)^{3/2}\bigg],
\end{equation}
where $C' = C[\ln(\mu{\bar t}^2\eta^2/CT_{\rm c})^{-5/6}]$. Since
the effect on $t_{\rm sp}$ only arises at the level of `$\ln\ln$'
terms, $C'\approx C$ is a good estimation in all that follows.
Since this choice underestimates $t_{\rm sp}$ it only strengthens our
results that $t_{\rm sp}>t_D$.

\section{The diffusion coefficient}
\label{dc-appendix}

In this Appendix we show how to obtain the diffussion coefficient Eq.(\ref{dqsq})
(for the quadratic coupling) from the reduced propagator. Following the same techniques
used for the quantum Brownian motion to obtain the master equation we must
compute the time derivative of the propagator $J_{\rm r}$ (Eq.(\ref{prosadle})), and eliminate
the dependence on the initial field configurations $\phi^\pm_{\rm i}$ that enters through
the classical solutions $\phi_{\rm cl}^\pm$. This can be easily done
using the propagator identities of Eq.(\ref{rel1}). Thus, using Eq.(\ref{us})
and identities (\ref{rel1}), we can write the initial field
configuration $\phi_{\rm i}^\pm$, in terms of the final configurations
($\phi_{\rm f}^\pm$), final field derivatives ($\partial_{\phi^\pm_{\rm f}}$),
and the functions of time $u_{\rm i}(s,t)$,

\begin{equation}
 (\phi_{\rm i}^{+} - \phi_{\rm i}^{-}) ~ u_1(s,t)~ J_0 = -
{{\dot u}_2(t,t)\over{{\dot u}_1(t,t)}} ~ u_1(s,t) ~ (
\phi_{\rm f}^{+} - \phi_{\rm f}^{-}) ~ J_0 + ...,
\label{cond}\end{equation}
neglecting terms proportional to derivatives respect to the final
field configuration which do not contribute to normal diffusion.

The temporal derivative is given by

\begin{eqnarray}i\hbar \partial_t J_r[&&\phi_{\rm f}^+,\phi_{\rm f}^-,t\vert
\phi_{\rm i}^+,\phi_{\rm i}^-,0] = \bigg\{h_{\rm ren}[\phi^+] - h_{\rm ren}[\phi^-] - i
g^2 {(\phi_{\rm f}^{+2} - \phi_{\rm f}^{-2})V\over{16}} \int_0^t ds ~ \Delta_2^{\rm cl}
(s)~ F(k_0;s,t)\nonumber \\ && + g^2
{(\phi_{\rm f}^{+2} + \phi_{\rm f}^{-2})V\over{16}} \int_0^t ds ~ \Delta_2^{\rm cl}
(s) ~ {\tilde K}_{\rm q}(k_0; s, t)
+ ... \bigg\}J_r[\phi_{\rm f}^+,\phi_{\rm f}^-,t\vert
\phi_{\rm i}^+,\phi_{\rm i}^-,0],\label{timeder} \end{eqnarray}
where $ {\tilde K}_{\rm q}(k_0; s, t)$ is the Fourier transform of the
dissipation kernel. The ellipsis denotes other terms coming from the time
derivative which do not contribute to diffusion.

Diffusive effects are associated with terms proportional
to $(\phi_{\rm f}^{+2} - \phi_{\rm f}^{-2})^2$ in the master
equation. Using Eq.(\ref{cond}) we remove initial conditions 
from Eq.(\ref{timeder}), and
looking only those terms proportional to ${\Delta_2^{\rm f}}^2$, we get
the master equation Eq.(\ref{master}). The diffusion coefficient
$D(k_0, t)$ comes from the noise contribution to the influence
functional and it is the time dependent coefficient that multiplies 
$\Delta_2^{\rm f2}$. Thus, for the quadratic coupling example, we find

\begin{equation} D_{\rm qu}(k_0, t) = \int_0^t ds ~ u(s,t) ~ F(k_0;s,t),
\end{equation}
where

$$u(s,t) = \left[u_2(s,t) - {{\dot u}_2(t,t)\over{{\dot u}_1(t,t)}}
u_1(s,t)\right]^2.$$

For different couplings between system and environment one should follow the
same procedure shown in this Appendix.

\end{document}